\documentclass[fleqn,usenatbib]{mnras}
\usepackage{newtxtext,newtxmath}
\usepackage[T1]{fontenc}
\usepackage{ae,aecompl,graphicx,amsmath,comment,epsfig,psfrag}

\title[FRB energy distribution]{Modelling the energy distribution in CHIME/FRB Catalog-1}

\author[Bhattacharyya et al.] {Siddhartha Bhattacharyya$^1$\thanks{sid.phy.in@gmail.com}, Somnath Bharadwaj$^1$, Himanshu Tiwari$^2$ and Suman Majumdar$^{3,4}$\\ 
$^1$Department  of Physics, Indian Institute of Technology, Kharagpur, India \\
$^2$International Centre for Radio Astronomy Research, Curtin University, Bentley, WA 6102, Australia\\
$^3$Department of Astronomy, Astrophysics and Space Engineering, Indian Institute of Technology, Indore, India\\
$^4$Department of Physics, Blackett Laboratory, Imperial College, London SW7 2AZ, U.K.}
\begin{document}
\label{firstpage}
\pagerange{\pageref{firstpage}--\pageref{lastpage}}
\maketitle

\begin{abstract}
We characterize the intrinsic properties of any FRB using its redshift $z$, spectral index $\alpha$ and energy $E_{33}$ in units of $10^{33} \, {\rm J}$ emitted across   $2128 - 2848\; {\rm MHz}$ in the FRB's rest frame. Provided that $z$ is inferred from the measured extra-galactic dispersion measure $DM_{\rm Ex}$,  the fluence $F$ of the observed event defines a track in  $(\alpha,E_{33})$ space which we refer to as the `energy track'. Here we consider the energy tracks for a sample of $254$ non-repeating low dispersion measure  FRBs from the CHIME/FRB Catalog-1, and use these to determine $n(E_{33} \mid \alpha)$ the conditional energy distribution i.e. the number of FRBs in the interval $\Delta E_{33}$ given a value of $\alpha$.  Considering $-10 \le \alpha \le 10$, we find that the entire energy scale shifts to higher energy values as $\alpha$ is increased. For all values of $\alpha$, we can identify two distinct energy ranges indicating that there are possibly two distinct FRB populations. At high energies, the distribution is well fitted by a modified Schechter function whose slope and characteristic energy both increase with $\alpha$. At low energies, the number of FRBs are in excess of the predictions of the modified Schechter function indicating that we may have a distinctly different population of low-energy FRBs.  We have checked that our main findings are reasonably robust to the assumptions regarding the Galactic Halo and Host galaxy contributions to the dispersion measure.  
\end{abstract}


\section{Introduction}
Fast Radio Bursts (FRBs) are short duration ($\sim {\rm ms}$), high energetic ($\sim 10^{32}-10^{36}\,{\rm J}$) dispersed radio pulses first detected at Parkes telescope \citep{lorimer07}. Till now, more than $800$ FRBs have been reported. The online catalog of these reported event can be found here\footnote{https://www.wis-tns.org//}. These reported FRBs can apparently be divided into two distinct classes ($1$) repeating FRBs and ($2$) non-repeating FRBs. 
 
The first repeating FRB event (FRB$121102$) was reported by Arecibo radio telescope \citep{spitler14}, and to-date there has been more than $30$ repeating bursts are reported from the same source (FRB$121102$) with identical values of its dispersion measure ($DM$). Other than Arecibo's detection, CHIME \citep{rafiei21} and VLA \citep{law20} have also reported the detection of repeating FRBs.   

On the other hand, a significant number of non-repeating FRBs $> 600$ have been reported at various instrument such as CHIME \citep{rafiei21}, Parkes \citep{price18a}, ASKAP \citep{bhandari20}, UTMOST \citep{gupta20}, FAST \citep{zhu20}, GBT \citep{parent20}, VLA \citep{law20}, DSA$-10$ \citep{ravi19}, Effelsberg \citep{marcote20}, SRT \citep{pilia20}, Apertif \citep{vanleeuwen22}, MeerKAT \citep{rajwade22}, etc. It is not clear whether the repeating and non-repeating FRBs are events drawn from the same parent population or not. Given the currently available data and the present understanding of FRBs, it is quite possible that the non-repeating FRBs form a separate population distinct from the repeating FRBs \citep{caleb18N,palaniswamy18,lu19}. The work reported here is entirely restricted to the non-repeating FRBs.

Several models have been proposed for the physical origin of the FRB emission. The reader is referred to \cite{zhang20a} and \cite{platts19} for a quick review on the proposed FRB progenitor models.  Considering observations, the intrinsic properties of the FRBs like the spectral index $\alpha$ and the energy distribution are rather poorly constrained.   \cite{macquart19} have analysed  a sample of $23$ FRBs detected at ASKAP to infer  a mean value of $\alpha=-1.5^{+0.2}_{-0.3}$,   whereas  \citet{houben19} have utilised the fact that  FRB $121102$ was not simultaneously detected  at $1.4\,{\rm GHz}$ and $150\,{\rm MHz}$  to proposed a lower limit $\alpha>-1.2\pm-0.4$.    \citet{james21} have modelled the FRB energy distribution using a simple power law, and  \citet{2022MNRAS.509.4775J} have used the FRBs detected at Parkes and ASKAP to constrain the exponent for the energy distribution to a value $-1.6^{+0.11}_{-0.12}$.  Various  studies \citep{macquart19,houben19,james21}  have each considered a specific telescope for which they have estimated the spectral index $\alpha$ of the  FRB emission.

An earlier work \citet{bera16}   has  modelled the FRB  population,  and used this to make predictions for FRB detection for different telescopes. In a subsequent work  \citet{bhattacharyya21}  have used the two-dimensional  Kolmogorov-Smirnov (KS)  test to compare the FRBs observed at Parkes, ASKAP, CHIME and UTMOST with simulated  predictions for different FRB population models. The parameter range $\alpha>4$ and $\overline{E}_{33}>60$ was found to be  ruled out with $95\%$ confidence, here  $\overline{E}_{33}$ is the mean energy  of the FRBs population  in units of  $10^{33}\,{\rm J}$ across $2128 - 2848\; {\rm MHz}$ in the FRB rest frame.  In a recent work \citep{bhattacharyya22} we have used  a sample of $82$ non-repeating FRBs detected at Parkes, ASKAP, CHIME and UTMOST to perform  a maximum likelihood analysis to determine the FRB population model parameters  which best fits this data. We have obtained  the best fit parameter values $\alpha=-1.53^{+0.29}_{-0.19}$, $\overline{E}_{33}=1.55^{+0.26}_{-0.22}$  and  $\gamma=0.77\pm 0.24$, where   $\gamma$ is the exponent of the Schechter  function energy distribution of  the FRBs.

The CHIME telescope\footnote{https://www.chime-frb.ca/home} has recently released the CHIME/FRB catalog $1$ \citep{amiri21} which has reported the detection of $535$ FRBs from $492$ different sources among which $462$ FRBs are apparently non-repeating. Several studies \citep{amiri21,rafiei21,chawla22,pleunis21,josephy21} have shown that there are  at least two distinct classes of FRBs in this catalog, \textit{viz.} ($1$)  where the observed $DM$ values lie within the range $100-500\,{\rm pc\,cm^{-3}}$ and ($2$)  where the observed $DM$ values are greater than $500\,{\rm pc\,cm^{-3}}$. It has  been proposed \citep{cui22}  that the  low $DM$  and  high $DM$  FRBs may originate from  two different kinds of sources,  or the sources may be the same with different environmental conditions accounting for the large $DM$ values. In the present work we have analyzed the energy distribution of the  non-repeating low $DM$  FRBs in the CHIME/FRB catalog $1$. A brief outline of this paper follows.  Section~\ref{sec:2} presents the methodology, while the results are presented in Section~\ref{sec:3} and we present our conclusions in Section~\ref{sec:4}.
In Appendix~\ref{sec:appendix}  we have analyzed how much  our results vary if we change the values of  $DM_{\rm Halo}$ and $DM_{\rm Host}$.

\section{Methodology}\label{sec:2}
\begin{figure}
\centering
\includegraphics[scale=0.65]{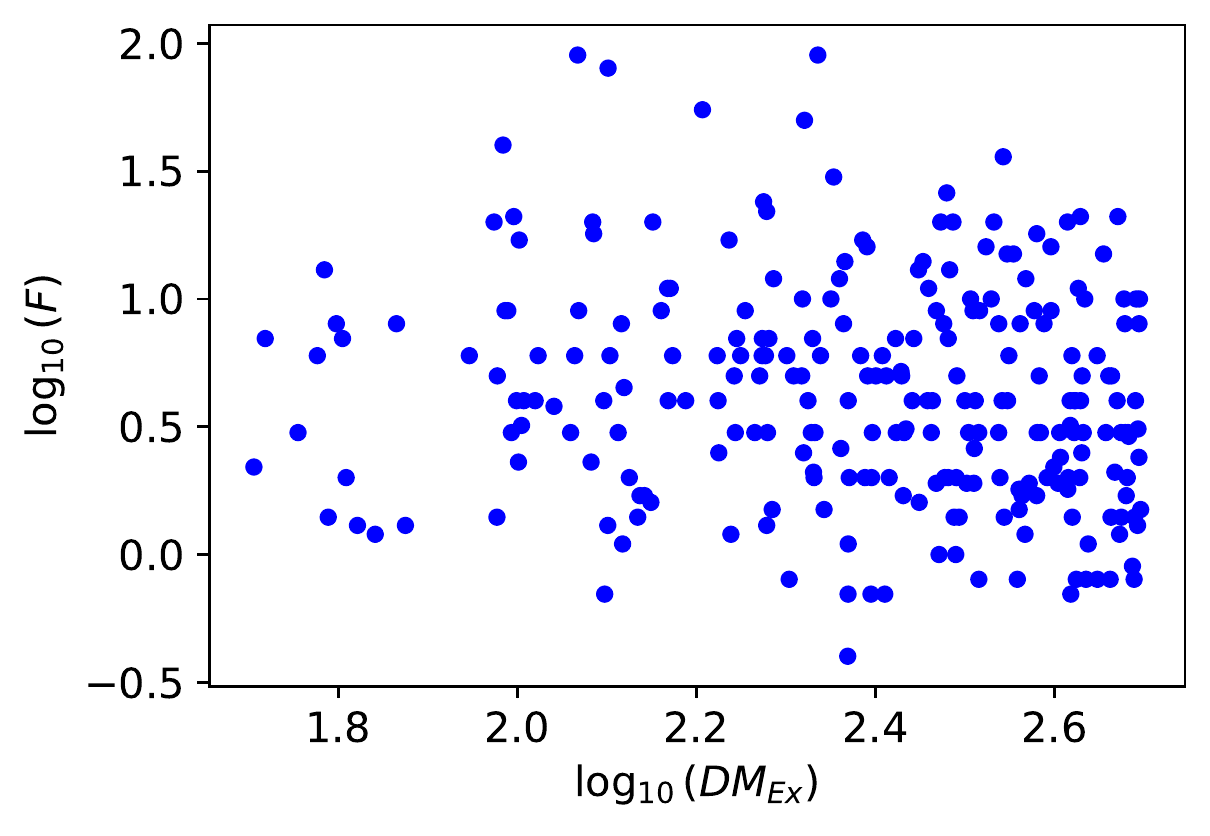}
\caption{The distribution of $DM_{\rm Ex}$ the extragalactic dispersion measure and $F$ the observed fluence of the $254$ non-repeating FRBs considered here. Using  a fixed value of $DM_{\rm Host}=50\,{\rm pc\,cm^{-3}}$, the redshift range of the these FRBs is $0.001\leq z\leq 0.46$.} 
\label{fig:observed-frbs}
\end{figure}

An observed FRB is characterized by its dispersion measure $DM$, fluence $F$ and pulse width $w$. For our analysis we consider $254$ non-repeating FRBs detected at CHIME \citep{amiri21} for which the extragalactic contribution to dispersion measure ($DM_{\rm Ex}$) lie within the range $50-500\,{\rm pc\,cm^{-3}}$. The $DM_{\rm Ex}$ for each FRB is estimated using the equation 
\begin{equation}
    DM_{\rm Ex}=DM_{\rm obs}-DM_{\rm MW}-DM_{\rm Halo}
    \label{eq:dmex-obs}
\end{equation}
where, $DM_{\rm obs}$ is the observed dispersion measure and $DM_{\rm MW}$ is the Milky Way contribution to $DM_{\rm obs}$  which is calculated using the NE$2001$ model \citep{cordes03}. $DM_{\rm Halo}$ is the Galactic halo contribution to $DM_{\rm obs}$, and we assume a fixed value $DM_{\rm Halo}=50\,{\rm pc\,cm^{-3}}$ \citep{prochaska19a} for all the FRBs considered here.
Figure \ref{fig:observed-frbs} shows the distribution of $DM_{\rm Ex}$ and $F$ of the $254$ non-repeating FRBs considered for this analysis. The $DM_{\rm Ex}$ and $F$ values of these FRBs lie within the range $50.8-496.8\,{\rm pc\,cm^{-3}}$ and $0.4-90\,{\rm Jy\,ms}$ respectively. We  do not see  any noticeable  correlation between $DM_{\rm Ex}$ and $F$ for  the FRBs considered here. Further, the dispersion of $F$ values also appears to be  independent of $DM_{\rm Ex}$.

There are no independent redshift measurements for the FRBs considered here, and we have used the extragalactic dispersion measure $DM_{\rm Ex}$  to infer the redshifts $(z)$ and hence the cosmological distances of the FRBs. Considering a homogeneous and completely ionized intergalactic medium (IGM), the $DM_{\rm Ex}$ can be modelled as 
\begin{equation}
    DM_{\rm Ex}(z)=940.51\int_0^z\,dz'\,\frac{1+z'}{\sqrt{\Omega_m\,(1+z')^3+\Omega_{\Lambda}}}+\frac{DM_{\rm Host}}{1+z}
    \label{eq:dmex}
\end{equation}
where the first term is the contribution from the inter-galactic medium (IGM) and the second term is the contribution from the host galaxy of an FRB. The unknown $DM_{\rm Host}$ is not well constrained for most of the unlocalized FRBs to date. Further,  $DM_{\rm Host}$ depends on the intersection  of the line of sight to  the  FRB with the disk of the host galaxy.  For the present analysis we mainly consider two scenarios for $DM_{\rm Host}$, (1) a fixed value $DM_{\rm Host}=50\,{\rm pc\,cm^{-3}}$ for all the FRBs and (2) a random value of $DM_{\rm Host}$ drawn from a log-normal distribution with mean $\log(50)$ and variance $0.5$. This two scenarios are denoted here as $DM_{50}$ and $DM_{\rm Rand50}$ respectively.  In addition to this, we have also repeated the entire analysis for several other choices of  $DM_{\rm Halo}$ and  $DM_{\rm Host}$ for which the results are presented in the Appendix.

The aim of this analysis is to model the energy distribution of the sample of  $254$ FRBs detected at CHIME. The FRB population model used here is presented in \cite{bera16},  and the reader is referred there for details. We have modelled the  intrinsic properties of an FRB using its spectral index $\alpha$, energy $E$ and intrinsic pulse width $w_i$. The pulse width does not figure in the present work, and we do not consider this here.  We first consider $E_{\nu}$ the  specific energy of an FRB which is  defined as the energy emitted in the frequency  interval $d \nu$ centred at the frequency $\nu$. 
We model this as 
\begin{equation}
E_{\nu}=E\,\phi(\nu)
\label{eq:a1}
\end{equation} 
where $E$ is the energy emitted in the frequency interval $\nu_a=2128\,{\rm MHz}$ to $\nu_b=2848\,{\rm MHz}$ in  the rest frame of the source, and  $\phi(\nu)$ is the emission profile which we have  modelled as a power law $\phi(\nu)\propto\nu^{\alpha}$.  We have assumed that all the FRBs have the same value of the spectral index $\alpha$. It is also convenient to introduce the average emission profile $\overline{\phi}(z,\alpha)$ which refers to a situation where an FRB at redshift $z$ is observed using a telescope which has a frequency coverage of $\nu_1$ to $\nu_2$. 
We then have  
 \begin{equation}
    \overline{\phi}(z,\alpha)=\left(\frac{\nu_2^{1+\alpha}-\nu_1^{1+\alpha}}{\nu_b^{1+\alpha}-\nu_a^{1+\alpha}}\right)\,\frac{(1+z)^{\alpha}}{\nu_2-\nu_1}
    \label{eq:phibar}
\end{equation}
which is the average emission profile at the rest frame of the source. Considering observations with CHIME, we have used   $\nu_1=400\,{\rm MHz}$ to  $\nu_2=800\,{\rm MHz}$ for the subsequent analysis.

For an FRB with energy $E$ and spectral index $\alpha$ located at redshift $z$, the fluence $F$ of  the observed event is predicted to be 
\begin{equation}
    F=\frac{E\,\overline{\phi}(z,\alpha)\,B(\vec{\theta})}{4\pi r^2(z)}
    \label{eq:fluence}
\end{equation}
where  $B(\vec{\theta})$ is the beam pattern of the telescope at the angular position $ \vec{\theta}$ of the event, and $r(z)$ is the comoving distance to  the source which we have  estimated using the flat $\Lambda CDM$ cosmology \citep{planck20}. The FRBs in our sample have all been detected very close to the beam center $(\mid \vec{\theta} \mid \le 0.05)$ for which $B(\vec{\theta}) \approx 1$  \citep{amiri21}, and the redshifts have all been inferred from $DM_{\rm Ex}$ using  eq.~(\ref{eq:dmex}). Finally we can use eq.~(\ref{eq:fluence}) to predict the observed fluence $F$ for any  FRB in our sample  provided we know two intrinsic properties of the FRB namely its energy $E$ and spectral index $\alpha$. Conversely, the observed fluence imposes a relation between  $E$ and $\alpha$  which we can visualise as a track  in the $\alpha-E$ plane.  We refer to this as the "energy track" of the FRB. It is further convenient to express the energy in units of $10^{33} \, {\rm J}$ which we denote as $E_{33}$.

\begin{table}
\caption{The $DM_{\rm Ex}$ and $F$ values of the four specific FRBs detected at CHIME.}
    \centering
    \begin{tabular}{cccc}
    \hline
    FRB & $DM_{\rm Ex}\,({\rm pc\,cm^{-3}})$ & $F\,({\rm Jy\,ms})$ & Comment \\
    \hline
    FRB$1$ & $50.8$ & $2.2$ & FRB with minimum $DM_{\rm Ex}$ \\
    FRB$2$ & $496.8$ & $1.5$ & FRB with maximum $DM_{\rm Ex}$ \\
    FRB$3$ & $233.9$ & $0.4$ & FRB with minimum $F$ \\
    FRB$4$ & $216.6$ & $90.0$ & FRB with maximum $F$ \\
    \hline
    \end{tabular}
    \label{tab:4FRBs}
\end{table}

\begin{figure}
\centering
\includegraphics[scale=0.55]{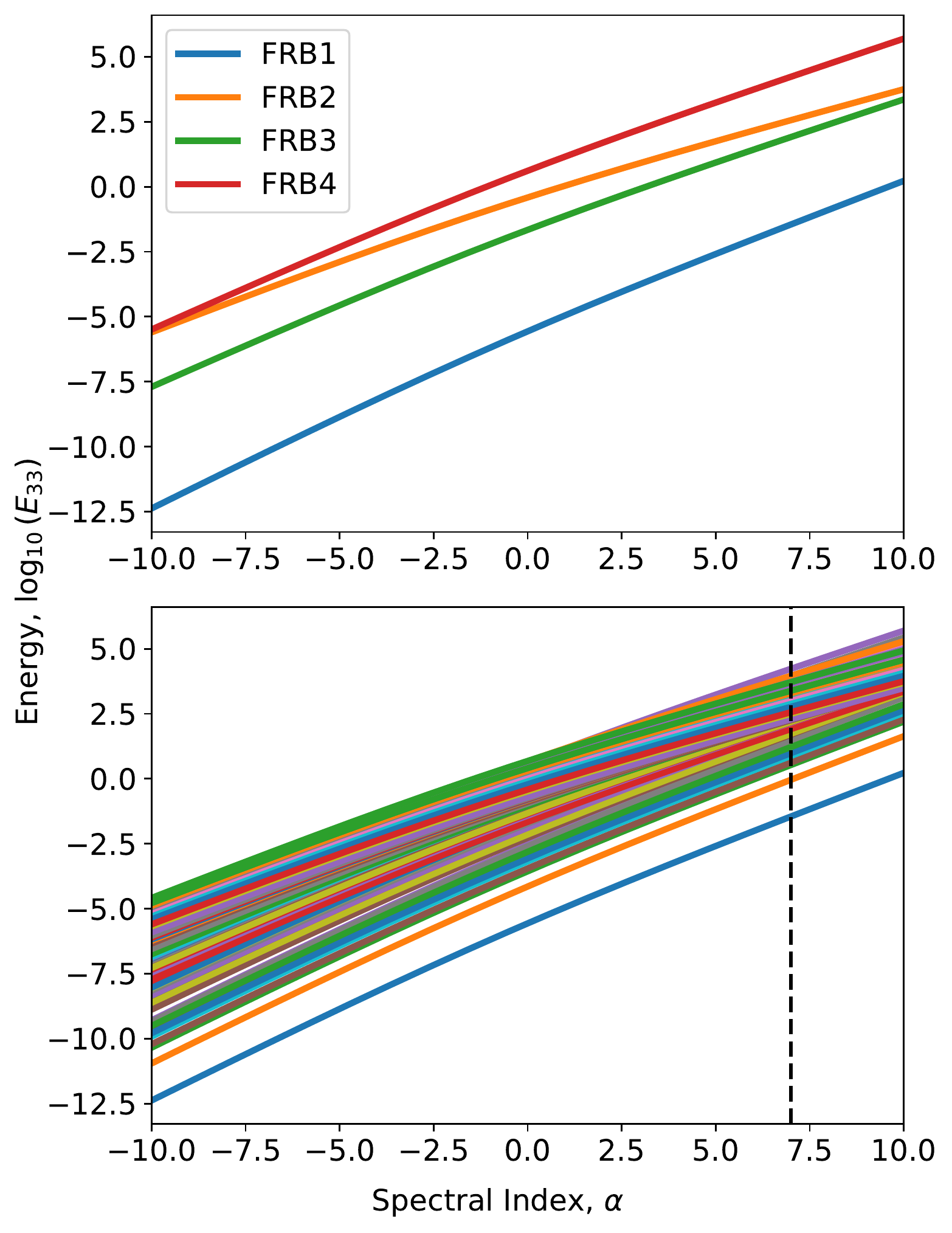}
\caption{The energy tracks i.e. the variation of the intrinsic energy $E_{33}$ with the spectral index $\alpha$ for the four specific FRBs (upper panel) and for all $254$ FRBs (lower panel) considered for this analysis. The vertical line in the lower panel corresponds to $\alpha=7$.}
\label{fig:energy-track}
\end{figure}

The upper panel of Figure~\ref{fig:energy-track} shows the energy tracks corresponding to four of the extreme FRBs in our sample for which  $DM_{\rm Ex}$ and $F$ are tabulated in Table~\ref{tab:4FRBs}.  Considering the two observed quantities $DM_{\rm Ex}$ and $F$, the shape of the track depends on $DM_{\rm Ex}$ (or equivalently $z$) through $\overline{\phi}(z,\alpha)$ whereas  the  track  just scales up or  down if the value of $F$ is changed. 
The term $\overline{\phi}(z,\alpha)$  accounts for the fact that we are actually probing a different frequency range for FRBs at different redshifts. 
The shape of $\overline{\phi}(z,\alpha)$  as a function of $\alpha$ changes with $z$. The FRBs in our sample  are all in the range $z<1$ for which the energy tracks are all found to all have a positive slope.  Comparing the tracks for FRB1 and FRB2, which have the highest and lowest $z$ respectively, we see that the track gets flatter as $z$ is increased. In fact, the slope reverses at $z\sim 4$, however our sample does not extend to such high $z$.   The FRBs with the minimum and maximum  fluence (FRB3 and FRB4 respectively) are also shown. Comparing  the four FRB tracks, we see that the track for FRB1   lies  considerably below the other three tracks which appear to lie close together within  a band.  This suggests that FRB1, which is the nearest FRB, is intrinsically much fainter (lower $E_{33}$) compared   to the other three FRBs shown in the figure.

The lower panel of Figure~\ref{fig:energy-track} shows the energy tracks for all the $254$ FRBs considered for the present analysis. We see that most of the FRB tracks are concentrated in a band which runs nearly diagonally across the figure, and the band gets narrow in its vertical extent ($E_{33}$) as $\alpha$ is increased. The  tracks of a few of the faintest FRBs, particularly the nearest one (FRB1 of Table~\ref{tab:4FRBs}), appear to lie distinctly below the band which encompasses the majority of the FRBs.

\begin{figure}
\centering
\includegraphics[scale=0.65]{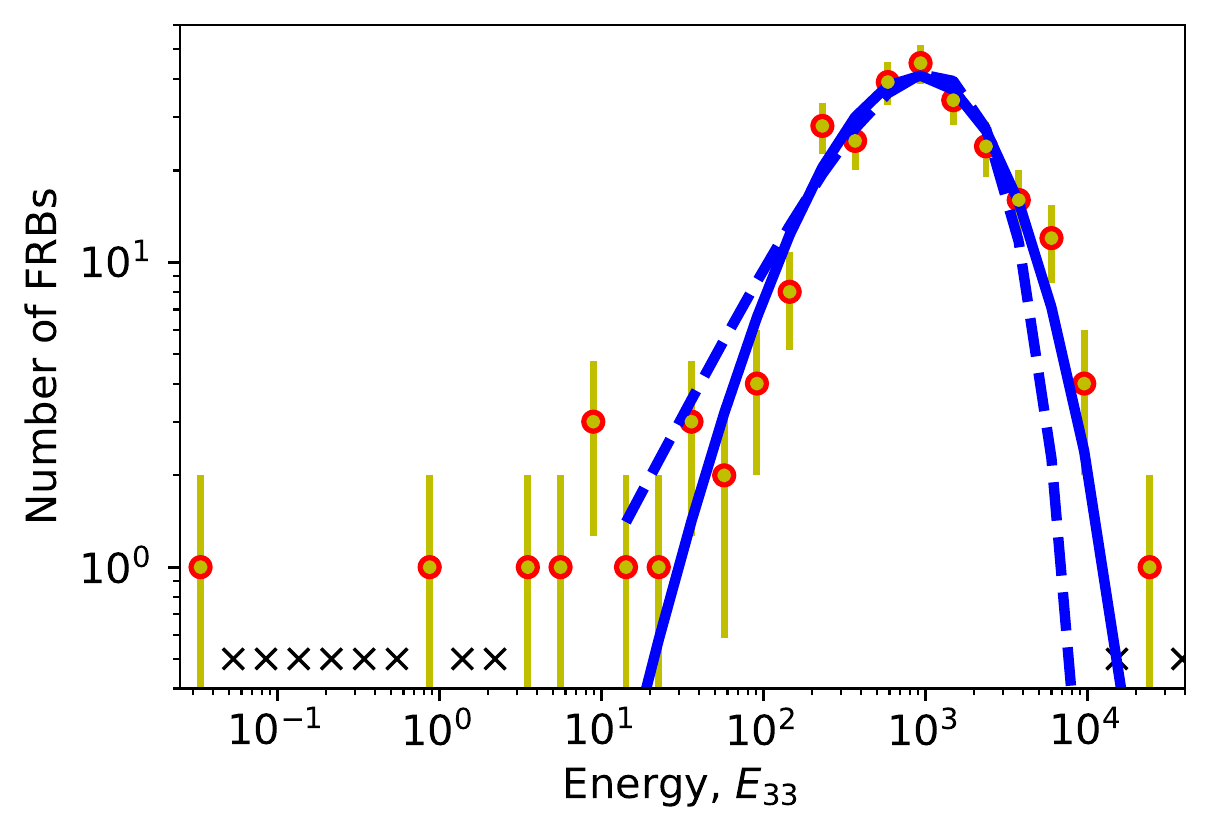}
\caption{The red circles show  $ N(E_{33} \mid \alpha)$ as a function of $E_{33}$ for $\alpha=7$, and  the yellow vertical lines show the $1 \sigma$  Poisson errors.  The black cross marks show the position of the empty bins. Considering eq.~(\ref{eq:schfunc}), the blue dashed  and solid lines show the predictions for the best fit with $\beta=1$ and $\beta=0.25$ respectively. }
\label{fig:dm50-curve-fit-single}
\end{figure}

The analysis presented till now has used the observed $ DM_{\rm Ex}$ and $F$ values of our  FRB sample to constrain the intrinsic FRB properties to a limited region of the two dimensional  $\alpha-E_{33}$ parameter space as shown in  Figure~\ref{fig:energy-track}.  However, for the subsequent analysis we focus on the one dimensional FRB energy $(E_{33})$ distribution. We have modelled this  using $ n(E_{33}\mid \alpha) $  the conditional energy distribution which is defined as  the numbers of FRBs in the energy interval $\Delta E_{33}$  centred around a value $E_{33}$  given a fixed  value of $\alpha$. Considering  the observational data, it is more convenient to  estimate $N(E_{33} \mid \alpha)$ which is the number of FRBs in bins of equal  logarithmic width  $\Delta (\log_{10} E_{33})=0.2$ given a fixed value of $\alpha$.  The two quantities introduced here are related as 
\begin{equation}
    N(E_{33} \mid \alpha)= 0.451  \,   E_{33}  \, \times   \,  n(E_{33}\mid \alpha) \,.
\label{eq:N}
\end{equation}
The procedure  for estimating $  N(E_{33} \mid \alpha)$   is illustrated in Figure~\ref{fig:energy-track} for $\alpha=7$  which is indicated by the vertical black dashed line.  The $E_{33}$ distribution  along this vertical line is divided into bins of  width $\Delta (\log_{10} E_{33})=0.2$ for which $ N(E_{33} \mid \alpha)$, the number of FRBs in each  bin,  is shown in Figure~\ref{fig:dm50-curve-fit-single}.  The black crosses  mark  the bins which do not contain any FRB {\it i.e.} $ N(E \mid \alpha)=0$, and the $1-\sigma$ error-bars correspond to $\sqrt{N(E \mid \alpha)}$   the  expected Poisson  fluctuations.

Considering Figure~\ref{fig:dm50-curve-fit-single}, we see that $N(E_{33} \mid \alpha) $ shows two distinctly different behaviour in two different $E_{33}$ ranges.   At low energies  in the range $E_{33}<10$ we find   several $E_{33}$ bins which have only a few $(\sim 1-3)$  FRBs each, whereas the remaining low $E_{33}$  bins  do not contain  any FRBs at all. We refer to this energy range as 'LELN' -  low $E_{33}$ low $N(E_{33} \mid \alpha) $.   We see that we have considerably higher  numbers of FRBS in each bin at higher energies in the range $E_{33} =10^2-10^4$. We refer to this energy range as  'HEHN' -  high $E_{33}$ high $N(E_{33} \mid \alpha) $.

We now focus on the FRBs in the HEHN range. Here  we see that $N(E_{33} \mid \alpha) $ exhibits a peak  at a characteristic energy $E_{33} \approx 10^{2}$. Further, $N(E_{33} \mid \alpha) $ appears to increases relatively gradually with $E_{33}$ in the range  $E_{33} < E_0$,  and it falls off  rapidly (nearly exponentially)  for $E_{33}>E_0$.  We find a similar behaviour also for other value of $\alpha$ (discussed later). Based on this  we have modelled  $n(E_{33}\mid \alpha)$  in the HEHN range as 
\begin{equation}
    n(E_{33} \mid \alpha )=K(\alpha) \left(\frac{E_{33}}{E_0(\alpha)}\right)^{\gamma(\alpha)}\exp\left[-\left(\frac{E_{33}}{E_0(\alpha)}\right)^{\beta}\right]
    \label{eq:schfunc}
\end{equation} 
where $\gamma(\alpha)$, $E_0(\alpha)$ and $K(\alpha)$ are three  parameters whose values are determined by  fitting  the observed  $N(E_{33} \mid \alpha)$ individually for each value of $\alpha$. We have not included $\beta$ as a parameter in the fit. However, we have tried out several discrete values of $\beta$ in order to assess which provides a good fit over the entire $\alpha$ range.  Considering $\beta=1$ we see that eq.~(\ref{eq:schfunc}) corresponds to the Schechter function  where $n(E_{33} \mid \alpha)$ and $N(E_{33} \mid \alpha)$ both decline  exponentially for $E_{33} >E_0$. 
Figure~\ref{fig:energy-track} shows the results for the  best fit Schechter function where we see that for $E_{33} >E_0$ the predicted  $N(E_{33} \mid \alpha)$  declines faster than the  actual  number of observed FRB events.  We find that the slower decline predicted for $\beta=0.25$ gives a better fit to the observed  $N(E_{33} \mid \alpha)$, particularly for $E_{33} >E_0$.  
Eq.~\ref{eq:schfunc} with $\beta=0.25$ (solid blue line) provides a better fit to the observed data in comparison to  $\beta=1$ (dashed blue line),  where the reduced $\chi^2$ have values $\chi^2_{\rm dof}=1.06$ and $2.72$ for $\beta=0.25$ and $1$ respectively.  Here we have identified  the first five non-empty  low $E_{33}$ bins as the LELN range which was  excluded from the fitting procedure. We refer to eq.~\ref{eq:schfunc} with $\beta=0.25$ as the "modified Schechter function".  Further, in the subsequent discussion we do not explicitly show $\alpha$ as an argument for the parameters $K, E_0$ and $\gamma$ for brevity of notation.  The methodology outlined here for $\alpha=7$,  was used for a range of $\alpha$ values for which the results are presented below.

\section{Results}\label{sec:3}

\begin{figure}
\centering
\includegraphics[scale=0.42]{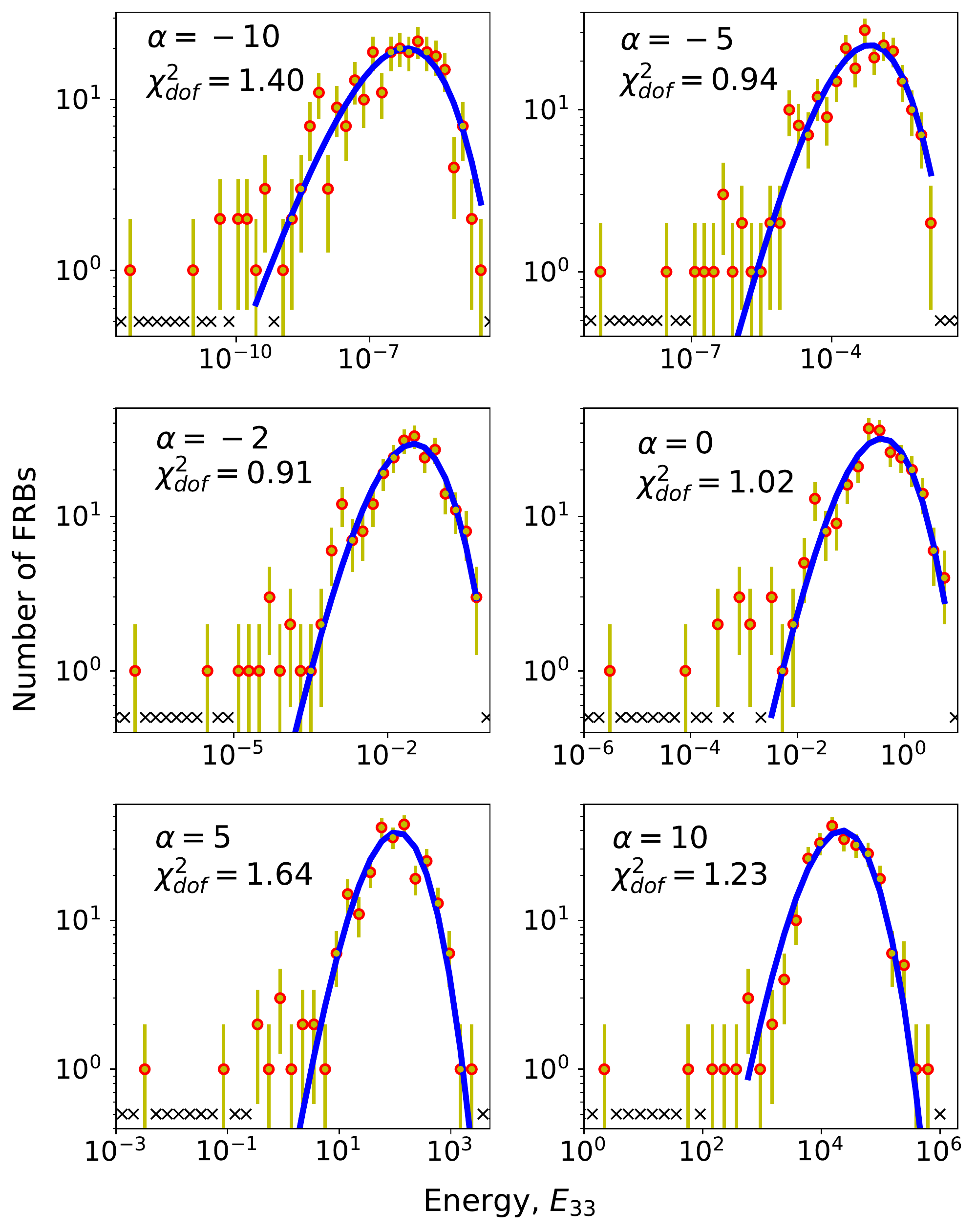}
\caption{Considering $DM_{50}$, the red circles show  $ N(E_{33} \mid \alpha)$ as a function of $E_{33}$ for different values of $\alpha$ as mentioned in the figure panels,  and  the yellow vertical lines show the $1 \sigma$  Poisson errors.  The black cross marks show the position of the empty bins. Considering eq.~(\ref{eq:schfunc}), the blue solid lines shows the predictions for the best fit modified Schechter function $(\beta=0.25)$ while the corresponding values of the reduced chi-squared  $\chi^2_{\rm dof}$ are mentioned in the respective figure panels.}
\label{fig:dm50-curve-fit}
\end{figure}

We have estimated $N(E_{33} \mid \alpha) $ considering  $\alpha$ in the range $-10 \ge \alpha \ge 10$.  The different panels of Figure~\ref{fig:dm50-curve-fit} show the results for $\alpha=-10$, $-5$, $-2$, $0$, $5$ and $10$ respectively, all considering $DM_{\rm Host}=50\,{\rm pc\,cm^{-3}}$ fixed, and we refer to this as $DM_{50}$.  We see that the behaviour is very similar for all values of $\alpha$, with the difference that the entire energy scale shifts to higher values as $\alpha$ is increased. In all cases we see that $N(E_{33} \mid \alpha) $ shows two distinctly different behaviour in two different energy ranges. For all values of $\alpha$, at low $E_{33}$ we find   several $E_{33}$ bins which have only a few $(\sim 1-3)$  FRBs each, whereas the remaining low $E_{33}$  bins  are empty. Here we have identified the first five  non-empty low $E_{33}$ bins as the LELN range.   
At higher $E_{33}$,  we see that $N(E_{33} \mid \alpha) $ initially increases with $E_{33}$, reaches a peak value and then declines. We refer to this as the HEHN range.   The energy distribution in the HEHN regime appears to be distinctly different from that in the LELN regime, and for the subsequent fitting we have excluded the FRBs in the LELN range. 
The best fit $N(E_{33} \mid \alpha) $  (blue line) and the $\chi^2_{\rm dof}$ values are shown in the different panels of Figure~\ref{fig:dm50-curve-fit}.  In all cases we find that the modified Schechter function provides a good fit to $N(E_{33} \mid \alpha) $  in the HEHN regime. 
We also see that in the LELN regime the observed $N(E_{33} \mid \alpha) $  values are considerably in excess of the predictions of the modified Schechter function, indicating that the low energy FRBs  have an energy distribution which is different from that of the high energy FRBs.

\begin{figure}
\centering
\includegraphics[scale=0.5]{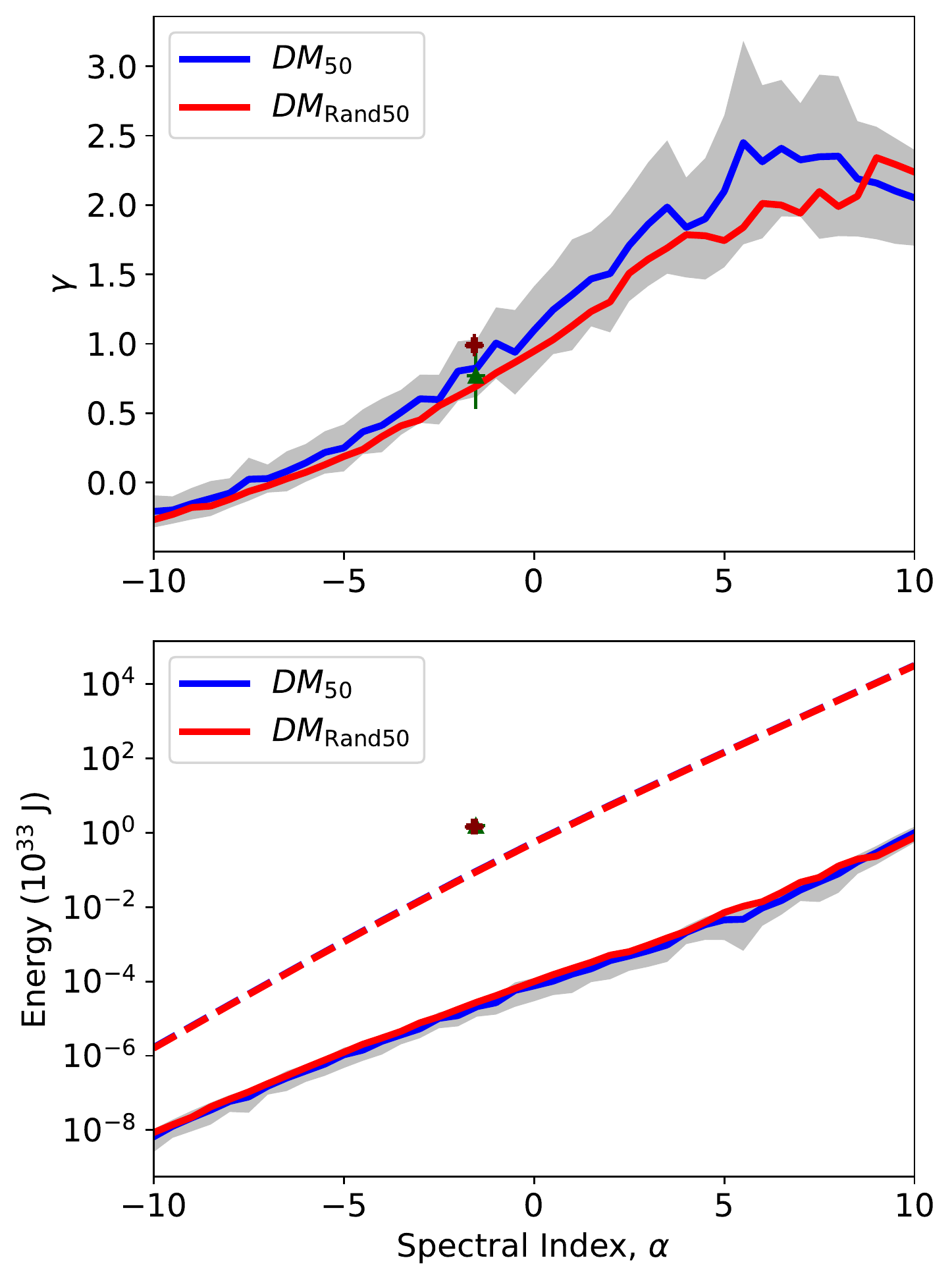}
\caption{The variation of fitting parameters $\gamma$ (upper panel) and $E_0$ (lower panel) with the spectral index $\alpha$. The blue and red lines correspond to $DM_{50}$ and $DM_{\rm Rand50}$ respectively.  The dashed blue and red lines in the lower panel show the variation of the average energy $\overline{E}_{33}$ with $\alpha$ for $DM_{50}$ and $DM_{\rm Rand50}$ respectively. The green triangle and brown plus symbols show the best fit parameter values $(\alpha,\gamma,\overline{E}_{33})$ obtained from our previous analysis \citep{bhattacharyya22} for $DM_{50}$ and $DM_{\rm Rand50}$ respectively.}
\label{fig:fitting-parameter}
\end{figure}

We next consider the variation of $\gamma$ and $E_0$ with $\alpha$ as shown in  Figure~\ref{fig:fitting-parameter}. We first consider the upper panel where the solid blue line shows the variation of $\gamma$ with $\alpha$. We see that $\gamma$ increases nearly linearly from $\gamma \approx -0.25$ at $\alpha=-10$ to  a maximum value of $\gamma \approx 2.43$ at $\alpha=6.5$ beyond which it declines slightly to $\gamma \approx 2$ at $\alpha=10$. 
Considering $E_0$ shown with the solid blue line in the lower panel, we see that $E_0$ increases exponentially with $\alpha$  with a value $E_0 \approx 10^{-8}$ for $\alpha=-10$ and $E_0 \approx 10^{0}$ for $\alpha=10$. We further see that  the $\alpha$   dependence of $\gamma$ and $E_0$ can respectively be  modelled    as 
\begin{equation}
\gamma \approx  0.17 \alpha + 1.1 \, \hspace{1cm}  {\rm and} \hspace{1cm}   E_0 \approx  10^{0.4 \alpha -4} \,. 
    \label{eq:e0}
\end{equation}
The lower panel also shows  the mean FRB energy $\overline{E}_{33}$ as a function of $\alpha$ (dashed line). We have evaluated this be considering the entire FRB sample (both LELN and HEHN). 

\begin{figure}
\centering
\includegraphics[scale=0.42]{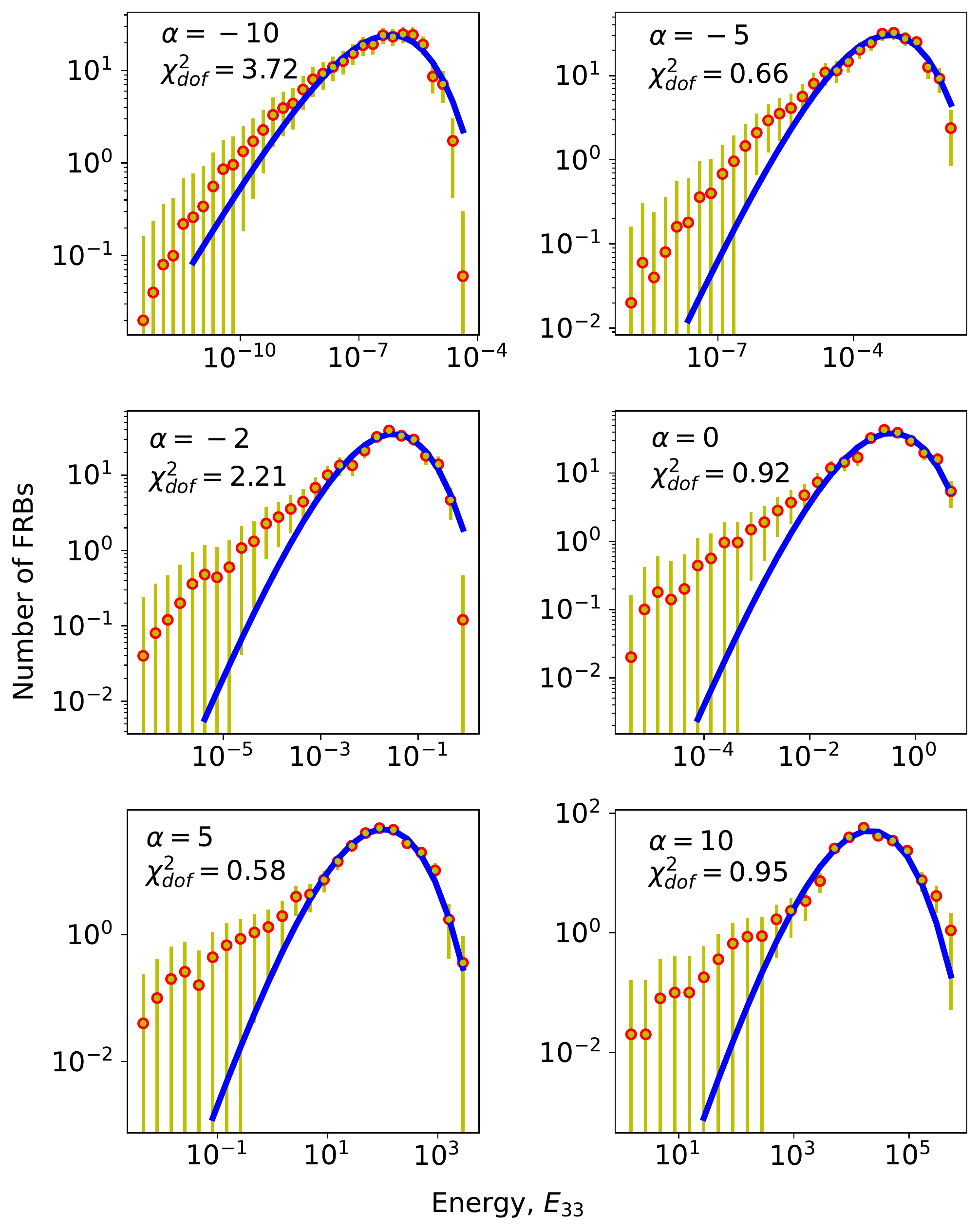}
\caption{Same as Figure~\ref{fig:dm50-curve-fit} for the $DM_{\rm Rand50}$}
\label{fig:dmrand-curve-fit}
\end{figure}

The entire analysis, until now, has assumed a fixed value  $DM_{\rm Host}=50\,{\rm pc\,cm^{-3}}$ for all the FRBs. We next consider the situation where  the $DM_{\rm Host}$ values have a  random distribution.  Here we have considered a situation where the $DM_{\rm Host}$ values are randomly drawn from a log-normal distribution with mean $\mu=\log(50)$ and standard deviation  $\sigma=0.5$, and we refer to this as $DM_{\rm Rand50}$.   We note that the inferred $z$ value depends on  $DM_{\rm Host}$  (eq.~\ref{eq:dmex}), and for each FRB it is necessary to impose an upper limit on the randomly generated  $DM_{\rm Host}$ in order to ensure that we can meaningfully estimate $z$. We have randomly generated $50$  values of $DM_{\rm Host}$ for each FRB which leads to $50$ energy tracks corresponding to each FRB. Considering the $254$ FRBs in our sample, we have averaged $50$ tracks for each FRB   to estimate  $N(E_{33} \mid \alpha) $ for which the results are shown in 
 Figure~\ref{fig:dmrand-curve-fit} for a few values of $\alpha$. We see that the results are quite similar to those  shown in Figure~\ref{fig:dm50-curve-fit} for $DM_{50}$, with the difference that there are no  empty bins  for $DM_{\rm Rand50}$. Here also we see a difference  between the FRB counts in the low and high energy ranges,  although  these differences are not as marked as as they were for $DM_{50}$. For each $\alpha$, the $E_{33}$ values used to demarcate the LELN and HEHN ranges for $DM_{\rm Rand50}$  were the same as those that we had used for  $DM_{50}$. 
 Considering the HEHN energy range,  for each value of $\alpha$ we have modelled $n(E_{33} \mid \alpha) $  using  a modified Schechter function.   The best fit $N(E_{33} \mid \alpha) $  (blue line) and the $\chi^2_{\rm dof}$ values are shown in the different panels of Figure~\ref{fig:dmrand-curve-fit}. Here also,  in all cases we find that the modified Schechter function provides a good fit to $N(E_{33} \mid \alpha) $  in the HEHN regime. 
 The $\alpha$ dependence of $\gamma$ and $E_0$ for $DM_{\rm Rand50}$ are also shown in Figure~\ref{fig:fitting-parameter}. We see that these are very similar to the results for $DM_{50}$, and we do not discuss them separately here.  
 
 In an earlier  work \citep{bhattacharyya22} we  have used  a sample of $82$ non-repeating FRBs detected at Parkes, ASKAP, CHIME and UTMOST to obtain the best fit parameter values $\alpha=-1.53^{+0.29}_{-0.19}$, $\overline{E}_{33}=1.55^{+0.26}_{-0.22}$  and  $\gamma=0.77\pm 0.24$. For comparison, we have shown these best fit values along with the $1 \sigma$ error bars in   Figure~\ref{fig:fitting-parameter}. Considering the  upper panel, we see that  the best fit values of $(\alpha,\gamma) $  obtained in our earlier work  are consistent with the results obtained in the present work. However, considering the lower panel, we see that the   best  fit values of $(\alpha, \overline{E})$  
 obtained in our earlier work are not consistent with the results obtained here.  The present work does not jointly constrain  $(\alpha, \overline{E}_{33}, \gamma)$, but instead  it provides  best fit values of $( \overline{E}_{33}, \gamma)$  given the value of $\alpha$. Considering $\alpha=-1.53$ which is  the best fit value from our earlier work, the present analysis gives $( \overline{E}_{33}, \gamma)=(0.173,0.625)$ for which the value of $\overline{E}_{33}$ falls below that  obtained in our earlier  work. However, it is necessary to note that a comparison is not straightforward as the approach used in the earlier work was quite different from that adopted here. The results from the earlier works depend on several input like (1.) the intrinsic FRB pulse width, (2.) the scattering model for pulse broadening in the IGM, and (3.) the $z$ dependence of the FRB event rate, neither of which effect the present analysis. Further, the earlier work had a different model for  $DM_{\rm Host}$. All of these factors can effect parameter estimation and  can possibly  account for the difference in the values of   $ \overline{E}_{33}$ .

The entire analysis, until now, has considered $DM_{\rm Halo}=50\,{\rm pc\,cm^{-3}}$ fixed, and two models for $DM_{\rm Host}$ namely $DM_{50}$ and $DM_{\rm Rand50}$ both of which have a reference value of $DM_{\rm Host}=50\,{\rm pc\,cm^{-3}}$. In order to check how sensitive our results are to these choices, we have redone the analysis considering other values of $DM_{\rm Halo}$  $(=40 \, {\rm and} \,  60 \,{\rm pc\,cm^{-3}})$ and    $DM_{\rm Host}$ $(=20, 40 \, {\rm and} \, 60 \, \,{\rm pc\,cm^{-3}})$. The results are presented in Appendix \ref{sec:appendix}. We find  that the  main findings  presented here are quite robust.  The changes in the values of $\gamma$ and $E_0$ are found to be  within $10 \%$  and $25 \%$ respectively.

\section{Conclusions}\label{sec:4}
We have modelled the intrinsic properties of an FRB using its energy $E_{33}$,  spectral index $\alpha$ and redshift $z$. These completely determine the observed fluence $F$  and extra-galactic dispersion measure $DM_{\rm Ex}$ provided other contributions to the $DM$ are known.  Here we have considered  $DM_{\rm Halo}=50\,{\rm pc\,cm^{-3}}$ and    $DM_{\rm Host}=50\,{\rm pc\,cm^{-3}}$, and used the measured  $DM_{\rm Ex}$ to infer $z$. The observed $F$ then restricts the intrinsic properties of the particular FRB to a one-dimensional track in the two-dimensional $\alpha-E_{33}$ parameter space. We refer to this as the energy track of the particular FRB. 

Here we have used  $254$  energy tracks corresponding to  a sample of FRBs detected at CHIME to  determine $N(E_{33} \mid \alpha)$ which is the number of  FRBs in bins of equal logarithmic  width $\Delta( \log_{10} ( E_{33}))=0.2$,  given a fixed value of $\alpha$. Considering $\alpha$ in the range  $-10 \le \alpha \le 10$,   we find that for all values of $\alpha$  the measured $N(E_{33} \mid \alpha)$  shows two distinctly different behaviour in different energy ranges.  The estimated $N(E_{33} \mid \alpha)$  has low values at low energies, and we refer to this as the LELN range. In contrast, at higher energies  the estimated $N(E_{33} \mid \alpha)$  increases with $E_{33}$, and  rises to a peak value beyond which it  declines. We refer to this as the HEHN range.In this paper we have modelled the FRB energy distribution in the HEHN range. 

We find that  in the HEHN range $n(E_{33}\mid \alpha) $,  the  conditional energy distribution, is well modelled by a modified Schechter function (eq.~\ref{eq:schfunc}). The values of the fitting parameters $\gamma$ and $E_0$ are found to increase with $\alpha$ (eq.~\ref{eq:e0}). The present analysis, however, does not pick out  a preferred value of $\alpha$. 

We can use the $n(E_{33}\mid \alpha) $ estimated in the HEHN range to predict the number of FRBs expected to be detected in the LELN range. We find that for all values of $\alpha$ these predictions are much below the the actual number of FRBs detected in the LELN range. This excess indicates that we  are possibly seeing a different population of FRBs at low energies. However, the number of events is currently not large enough to study the FRBs energy distribution in  the LELN range.  We hope to address this in future when more FRB detections become  available.  Finally,  we have checked that the findings presented in this paper do not change very much even if we consider different values of $DM_{\rm Halo}$ and    $DM_{\rm Host}$. 

In future, we hope to carry out  a combined   analysis of the FRBs detected at different telescopes and use this to simultaneously  constrain all three parameters $(\alpha,E_0,\gamma)$.

%\section*{Acknowledgement}

%\section*{Data availability}
%The data and codes underlying this article will be shared on reasonable request to the corresponding author.

\bibliographystyle{mnras}
\bibliography{reference}

\begin{thebibliography}{}
\makeatletter
\relax
\def\mn@urlcharsother{\let\do\@makeother \do\$\do\&\do\#\do\^\do\_\do\%\do\~}
\def\mn@doi{\begingroup\mn@urlcharsother \@ifnextchar [ {\mn@doi@}
  {\mn@doi@[]}}
\def\mn@doi@[#1]#2{\def\@tempa{#1}\ifx\@tempa\@empty \href
  {http://dx.doi.org/#2} {doi:#2}\else \href {http://dx.doi.org/#2} {#1}\fi
  \endgroup}
\def\mn@eprint#1#2{\mn@eprint@#1:#2::\@nil}
\def\mn@eprint@arXiv#1{\href {http://arxiv.org/abs/#1} {{\tt arXiv:#1}}}
\def\mn@eprint@dblp#1{\href {http://dblp.uni-trier.de/rec/bibtex/#1.xml}
  {dblp:#1}}
\def\mn@eprint@#1:#2:#3:#4\@nil{\def\@tempa {#1}\def\@tempb {#2}\def\@tempc
  {#3}\ifx \@tempc \@empty \let \@tempc \@tempb \let \@tempb \@tempa \fi \ifx
  \@tempb \@empty \def\@tempb {arXiv}\fi \@ifundefined
  {mn@eprint@\@tempb}{\@tempb:\@tempc}{\expandafter \expandafter \csname
  mn@eprint@\@tempb\endcsname \expandafter{\@tempc}}}

\bibitem[\protect\citeauthoryear{Aghanim, Akrami  et~al.}{Aghanim
  et~al.}{2020}]{planck20}
Aghanim N.,  Akrami Y.,   et~al., 2020, A\&A, 641, A12

\bibitem[\protect\citeauthoryear{Amiri, Andersen  et~al.}{Amiri
  et~al.}{2021}]{amiri21}
Amiri M.,  Andersen B.~C.,   et~al., 2021, arXiv preprint, 2106.04352

\bibitem[\protect\citeauthoryear{Bera, Bhattacharyya  et~al.}{Bera
  et~al.}{2016}]{bera16}
Bera A.,  Bhattacharyya S.,   et~al., 2016, MNRAS, 457, 2530

\bibitem[\protect\citeauthoryear{Bhandari, Sadler  et~al.}{Bhandari
  et~al.}{2020}]{bhandari20}
Bhandari S.,  Sadler E.~M.,   et~al., 2020, ApJ Letters, 895, L37

\bibitem[\protect\citeauthoryear{Bhattacharyya \& Bharadwaj}{Bhattacharyya \&
  Bharadwaj}{2021}]{bhattacharyya21}
Bhattacharyya S.,  Bharadwaj S.,  2021, MNRAS, 502, 904

\bibitem[\protect\citeauthoryear{Bhattacharyya, Tiwari  et~al.}{Bhattacharyya
  et~al.}{2022}]{bhattacharyya22}
Bhattacharyya S.,  Tiwari H.,   et~al., 2022, MNRAS: Letters, 513, L1

\bibitem[\protect\citeauthoryear{Caleb, Spitler  \& Stappers}{Caleb
  et~al.}{2018}]{caleb18N}
Caleb M.,  Spitler L.~G.,   Stappers B.~W.,  2018, Nature Astronomy, 2, 839

\bibitem[\protect\citeauthoryear{Chawla, Kaspi  et~al.}{Chawla
  et~al.}{2022}]{chawla22}
Chawla P.,  Kaspi V.~M.,   et~al., 2022, The Astrophysical Journal, 927, 35

\bibitem[\protect\citeauthoryear{Cordes \& Lazio}{Cordes \&
  Lazio}{2003}]{cordes03}
Cordes J.~M.,  Lazio T. J.~W.,  2003, arXiv preprint, astro-ph/0207156

\bibitem[\protect\citeauthoryear{Cui, Zhang  et~al.}{Cui et~al.}{2022}]{cui22}
Cui X.,  Zhang C.,   et~al., 2022, Astrophysics and Space Science, 367, 1

\bibitem[\protect\citeauthoryear{Gupta, Bailes  et~al.}{Gupta
  et~al.}{2020}]{gupta20}
Gupta V.,  Bailes M.,   et~al., 2020, The Astronomer's Telegram, 13788, 1

\bibitem[\protect\citeauthoryear{Houben, Spitler  et~al.}{Houben
  et~al.}{2019}]{houben19}
Houben L. J.~M.,  Spitler L.~G.,   et~al., 2019, A \& A, 623, A42

\bibitem[\protect\citeauthoryear{James, Prochaska  et~al.}{James
  et~al.}{2021}]{james21}
James C.~W.,  Prochaska J.~X.,   et~al., 2021, arXiv preprint, 2101.08005

\bibitem[\protect\citeauthoryear{{James}, {Prochaska}, {Macquart},
  {North-Hickey}, {Bannister}  \& {Dunning}}{{James}
  et~al.}{2022}]{2022MNRAS.509.4775J}
{James} C.~W.,  {Prochaska} J.~X.,  {Macquart} J.~P.,  {North-Hickey} F.~O.,
  {Bannister} K.~W.,   {Dunning} A.,  2022, \mn@doi [\mnras]
  {10.1093/mnras/stab3051}, \href
  {https://ui.adsabs.harvard.edu/abs/2022MNRAS.509.4775J} {509, 4775}

\bibitem[\protect\citeauthoryear{Josephy, Chawla  et~al.}{Josephy
  et~al.}{2021}]{josephy21}
Josephy A.,  Chawla P.,   et~al., 2021, The Astrophysical Journal, 923, 2

\bibitem[\protect\citeauthoryear{Law et~al.,}{Law et~al.}{2020}]{law20}
Law C.~J.,  et~al., 2020, The Astrophysical Journal, 899, 161

\bibitem[\protect\citeauthoryear{Lorimer, Bailes  et~al.}{Lorimer
  et~al.}{2007}]{lorimer07}
Lorimer D.~R.,  Bailes M.,   et~al., 2007, Science, 318, 777

\bibitem[\protect\citeauthoryear{Lu \& Piro}{Lu \& Piro}{2019}]{lu19}
Lu W.,  Piro A.~L.,  2019, ApJ, 883, 40

\bibitem[\protect\citeauthoryear{Macquart, Shannon  et~al.}{Macquart
  et~al.}{2019}]{macquart19}
Macquart J.~P.,  Shannon R.~M.,   et~al., 2019, ApJ Letters, 872, L19

\bibitem[\protect\citeauthoryear{Marcote et~al.,}{Marcote
  et~al.}{2020}]{marcote20}
Marcote B.,  et~al., 2020, Nature, 577, 190

\bibitem[\protect\citeauthoryear{Palaniswamy, Li  \& Zhang}{Palaniswamy
  et~al.}{2018}]{palaniswamy18}
Palaniswamy D.,  Li Y.,   Zhang B.,  2018, ApJ Letters, 854, L12

\bibitem[\protect\citeauthoryear{Parent et~al.,}{Parent
  et~al.}{2020}]{parent20}
Parent E.,  et~al., 2020, The Astrophysical Journal, 904, 92

\bibitem[\protect\citeauthoryear{Pilia et~al.,}{Pilia et~al.}{2020}]{pilia20}
Pilia M.,  et~al., 2020, The Astrophysical Journal Letters, 896, L40

\bibitem[\protect\citeauthoryear{Platts, Weltman  et~al.}{Platts
  et~al.}{2019}]{platts19}
Platts E.,  Weltman A.,   et~al., 2019, Physics Reports, 821, 1

\bibitem[\protect\citeauthoryear{Pleunis, Good  et~al.}{Pleunis
  et~al.}{2021}]{pleunis21}
Pleunis Z.,  Good D.~C.,   et~al., 2021, The Astrophysical Journal, 923, 1

\bibitem[\protect\citeauthoryear{Price et~al.,}{Price et~al.}{2018}]{price18a}
Price D.~C.,  et~al., 2018, The Astronomer's Telegram, 11376, 1

\bibitem[\protect\citeauthoryear{Prochaska \& Zheng}{Prochaska \&
  Zheng}{2019}]{prochaska19a}
Prochaska J.~X.,  Zheng Y.,  2019, MNRAS, 485, 648

\bibitem[\protect\citeauthoryear{Rafiei-Ravandi et~al.,}{Rafiei-Ravandi
  et~al.}{2021}]{rafiei21}
Rafiei-Ravandi M.,  et~al., 2021, The Astrophysical Journal, 922, 42

\bibitem[\protect\citeauthoryear{Rajwade, Bezuidenhout  et~al.}{Rajwade
  et~al.}{2022}]{rajwade22}
Rajwade K.~M.,  Bezuidenhout M.~C.,   et~al., 2022, arXiv preprint, 2205.14600

\bibitem[\protect\citeauthoryear{Ravi et~al.,}{Ravi et~al.}{2019}]{ravi19}
Ravi V.,  et~al., 2019, Nature, 572, 352

\bibitem[\protect\citeauthoryear{Spitler, Cordes  et~al.}{Spitler
  et~al.}{2014}]{spitler14}
Spitler L.~G.,  Cordes J.~M.,   et~al., 2014, ApJ, 790, 101

\bibitem[\protect\citeauthoryear{Zhang}{Zhang}{2020}]{zhang20a}
Zhang B.,  2020, Nature, 587, 45

\bibitem[\protect\citeauthoryear{Zhu et~al.,}{Zhu et~al.}{2020}]{zhu20}
Zhu W.,  et~al., 2020, The Astrophysical Journal Letters, 895, L6

\bibitem[\protect\citeauthoryear{van Leeuwen, Kooistra  et~al.}{van Leeuwen
  et~al.}{2022}]{vanleeuwen22}
van Leeuwen J.,  Kooistra E.,   et~al., 2022, arXiv preprint, 2205.12362

\makeatother
\end{thebibliography}

\appendix

\section{The effect of varying $DM_{\rm Halo}$ and $DM_{\rm Host}$}\label{sec:appendix}
\begin{figure*}
\centering
\includegraphics[scale=0.45]{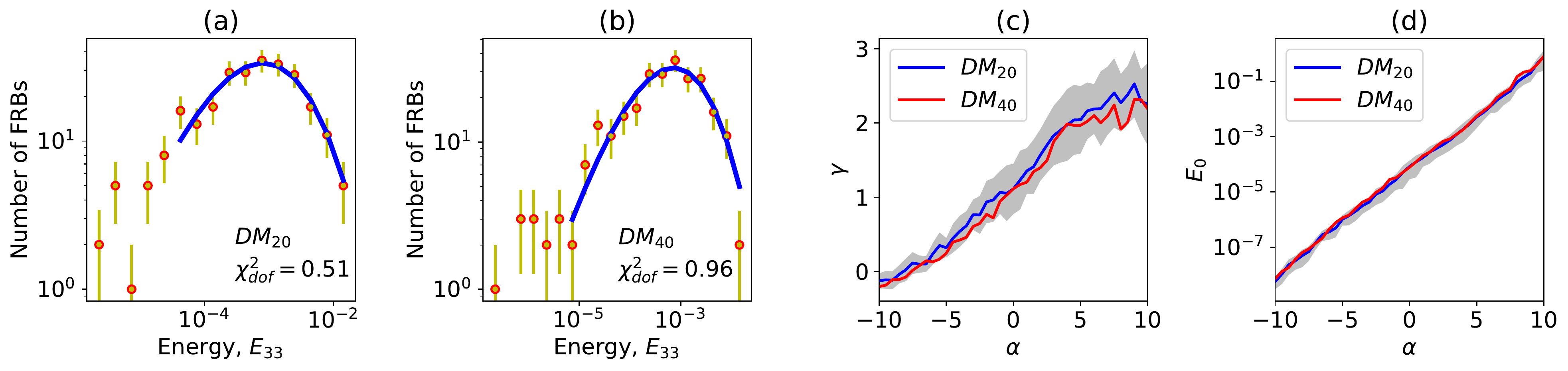}
\caption{Results for  two $DM_{\rm Host}$ models as indicated in the figure. Panels (a) and (b) show $N(E_{33} \mid \alpha)$ (with $1 \sigma$ errors) as a function of $E_{33}$ for $\alpha=-5$, while the solid line shows the predictions of the best fit modified Schechter function. The value of the reduced chi-square for the fit is indicated in the corresponding panels. Panels (c) and (d) respectively show the variation of the fitting parameters $\gamma$ and $E_0$ with $\alpha$.}
\label{fig:DMfix}
\end{figure*}

\begin{figure*}
\centering
\includegraphics[scale=0.45]{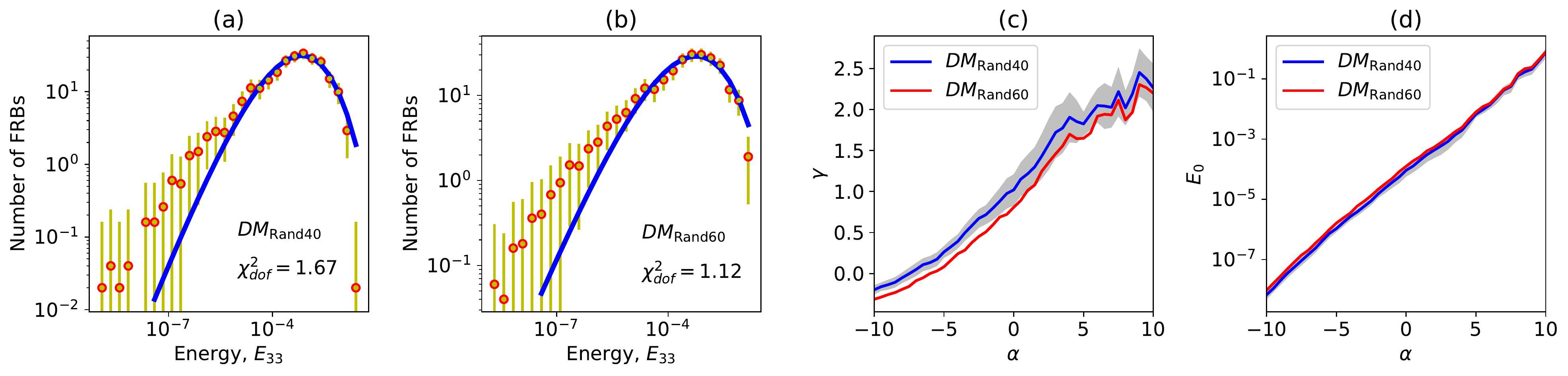}
\caption{Same as Figure \ref{fig:DMfix} but with  two different  $DM_{\rm Host}$ models as indicated in the figure.}
\label{fig:DMrand}
\end{figure*}

\begin{figure*}
\centering
\includegraphics[scale=0.45]{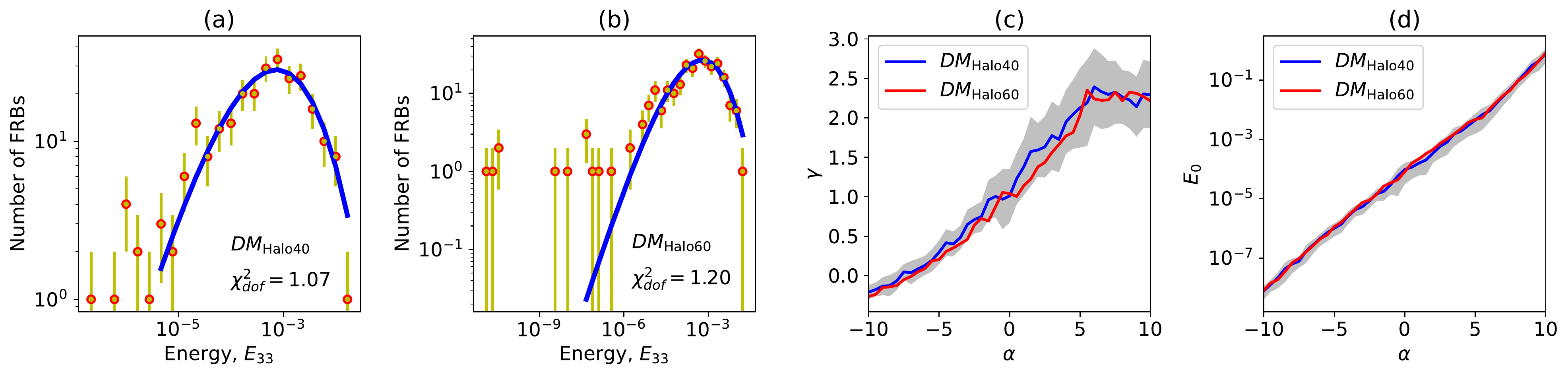}
\caption{Same as Figure \ref{fig:DMfix} but with   two different  $DM_{\rm Halo}$ models as indicated in the figure.}
\label{fig:DMhalo}
\end{figure*}

Here we have studied the effect of varying $DM_{\rm Halo}$ and $DM_{\rm Host}$. 
We first consider $DM_{\rm Halo}=50\,{\rm pc\,cm^{-3}}$ fixed, and vary the value of 
$DM_{\rm Host}$ to $20\,{\rm pc\,cm^{-3}}$ ($DM_{20}$) and $40\,{\rm pc\,cm^{-3}}$ ($DM_{40}$) for which  the results are shown in Figure~\ref{fig:DMfix}. 
It is not possible to consider a fixed $ DM_{\rm Host}>50\,{\rm pc\,cm^{-3}} $ as this would exceed $DM_{\rm Ex}$ for some of the FRBs in our sample.  We have next considered  random $DM_{\rm Host}$ values drawn from log-normal distributions with $(\mu,\sigma)=(\log(40),0.4)$ and $(\log(60),0.6)$ for which the results are shown in  Figure~\ref{fig:DMrand}.  
Finally, we have varied $DM_{\rm Halo}$ to $40 \, {\rm and} \, 60\,{\rm pc\,cm^{-3}}$ maintaining $DM_{\rm Host}=50\,{\rm pc\,cm^{-3}}$ fixed, for which the results are shown in Figure~\ref{fig:DMhalo}. 
In all of these figures  panels (a) and (b) show  $N(E_{33} \mid \alpha)$  for $\alpha=-5$. These panels also show the predictions of the best fit modified Schechter function. In all of the cases considered here, we find that the modified Schechter function provides a good fit to the observed FRB energy distribution  in the HEHN region. Panels (c) and (d) of all the figures show the  fitting parameters $\gamma$ and $E_0$ for different values of $\alpha$. Considering   $DM_{50}$ (Figure \ref{fig:fitting-parameter}) as the reference model,  and comparing the results with those in  Figure~\ref{fig:DMfix}-\ref{fig:DMhalo} we find that the changes in the values of $\gamma$ and $E_0$ are  within $10 \%$  and $25 \%$ respectively.

\label{lastpage}

\end{document}